\documentclass[prd,showpacs,twocolumn,preprintnumbers,amsmath,amssymb,nofootinbib,superscriptaddress]{revtex4-2}
\usepackage{epsfig}
\usepackage[utf8]{inputenc}
\usepackage{color}
\usepackage{epstopdf}
\usepackage{multirow}
\usepackage{verbatim}
\usepackage{ftnxtra}
\usepackage{slashed}
\usepackage{ulem}
\usepackage{subfigure}
\usepackage[colorlinks=true,linkcolor=blue,breaklinks=true,urlcolor=blue,citecolor=blue]{hyperref}

\newcommand{\be}{\begin{eqnarray}}
\newcommand{\ee}{\end{eqnarray}}

\newcommand{\uestc}{\affiliation{School of Physics, University of Electronic Science and
Technology of China, Chengdu 610054, China}}
\newcommand{\ucas}{\affiliation{School of Physical Sciences, University of Chinese Academy of Sciences (UCAS), Beijing 100049, China}}
\newcommand{\kek}{\affiliation{KEK Theory Center, Institute of Particle and Nuclear Studies (IPNS), High Energy Accelerator Research Organization (KEK), 1-1 Oho, Tsukuba, Ibaraki, 305-0801, Japan}}

\newcommand{\asrc}{\affiliation{Advanced Science Research Center, Japan Atomic Energy
Agency, Tokai, Ibaraki, 319-1195, Japan}}

\newcommand{\riken}{\affiliation{Nishina Center
for Accelerator-Based Science, RIKEN, Wako 351-0198, Japan}}

\newcommand{\pku}{\affiliation{School of Physics and Center of High Energy Physics,
Peking University, Beijing 100871,China}}

\begin{document}

\title{
New insight into the exotic states strongly coupled with the
$D\bar{D}^*$ from the $T^+_{cc}$}

\author{Guang-Juan Wang}\email{wgj@post.kek.jp}
\kek

\author{Zhi Yang}\email{zhiyang@uestc.edu.cn, corresponding author}
\uestc

\author{Jia-Jun Wu}\email{wujiajun@ucas.ac.cn, corresponding author}
\ucas

\author{Makoto Oka}\email{makoto.oka@riken.jp}
\asrc \riken

\author{Shi-Lin Zhu}\email{zhusl@pku.edu.cn}
\pku

\date{\today}

\begin{abstract}
We have investigated the internal structure of the open- and
hidden-charmed ($DD^*$/$\bar DD^*$) molecules in the unified
framework.
We first fit the experimental lineshape of the $T^+_{cc}$ state and
extract the $DD^*$ interaction, from which the $T^+_{cc}$ is assumed
to arise solely. Then we obtain the $D\bar{D}^*$ interaction by
charge conjugation.
Our results show that the $D\bar{D}^*$ interaction is attractive but
insufficient to form $X(3872)$ as a bound state.
Instead, its formation requires the crucial involvement of the
coupled channel effect between the $D\bar D^*$ and $c\bar c$
components, although the $c\bar c$ accounts for approximately $1\%$
only. Besides $X(3872)$, we have obtained a higher-energy state
around $3957.9$ MeV with a width of $16.7$ MeV, which may be a
potential candidate for the $X(3940)$.
In  $J^{PC}=1^{+-}$ sector, we have found two states related to the
iso-scalar $\tilde X(3872)$ and $h_c(2P)$, respectively.
Our combined study provides valuable insights into the nature of
these $DD^*$/$D\bar D^*$ exotic states.

\end{abstract}

\maketitle

\section{Introduction}
The exotic states, which are beyond the simple quark model pictures
of the quark-antiquark meson and three-quark baryon, have greatly
enriched the hadron spectroscopy.
Their inner structures remain a mystery and the investigations of
their dynamics have been an ongoing and central issue in the study
of nonperturbative Quantum Chromodynamics (QCD).

One of the most well-known exotic states is the $X(3872)$ with
$J^{PC}=1^{++}$ (also named as $\chi_{c1}(3872)$ in
RPP~\cite{ParticleDataGroup:2022pth}), which was first observed by
the Belle Collaboration in 2003~\cite{Belle:2003nnu} and
subsequently confirmed by various experimental
collaborations~\cite{CDF:2003cab,D0:2004zmu,BaBar:2004oro,LHCb:2011zzp,LHCb:2013kgk,CMS:2013fpt}.
It is located extremely near the $D^0\bar{D}^{*0}$
\footnote{Hereafter, we use the notion $D\bar D^*$ to represent
$D\bar D^*$+c.c..} threshold, with a mass difference of
$M_{X(3872)}-M_{D^0\bar{D}^{*0}}=0\pm0.18$ MeV.
The location is also not too far away from the $\chi_{c1}({2P})$
meson predicted by the Godfrey-Isgur (GI) relativized quark model
~\cite{Godfrey:1985xj}, with $M_{\chi_{c1}(2P)}-M_{X(3872)}=81.4$
MeV.
So far, there are different theoretical interpretations, such as the
conventional twisted  $\chi_{c1}(2P)$
charmonium~\cite{Barnes:2003vb,Kalashnikova:2010hv}, the compact
tetraquark state, the $ D^{{*}}\bar D$/$ D\bar D^{{*}}$
molecule~\cite{Tornqvist:2004qy,Liu:2008fh,Liu:2009qhy,Li:2012cs},
the mixture of the $c\bar c$ and  $ D^{{*}}\bar D$/$ D\bar D^{{*}}$
molecule \cite{Braaten:2003he, Kalashnikova:2005ui, Barnes:2007xu,
Ortega:2009hj, Li:2009ad, Baru:2010ww, Cincioglu:2016fkm,
Yamaguchi:2019vea}.
For more details, see Refs.~\cite{Chen:2016qju, Esposito:2016noz,
Brambilla:2019esw, Kalashnikova:2018vkv,Wu:2022ftm} for reviews.

The $X(3872)$ is challenging to understand due to its proximity to
the $D\bar D^*$ threshold and the intricate interactions between the
components $c\bar c$ and the $D\bar D^*$.
In traditional methods, the parameters governing the $D\bar D^*$
interaction are often estimated through phenomenological models,
leading to substantial uncertainties, especially for the
near-threshold states.
Within the molecular scenario, there are two possible fine-tuning
mechanisms for the $X(3872)$: either the accidental fine-tuning of
the parameters in the $D\bar D^*$ sector or an accidental
fine-tuning of a P-wave $c\bar c$ meson to the $D\bar D^*$
threshold~\cite{Braaten:2003he}.
Without constraints on these two parts, determining their roles
solely based on limited observables is practically impossible.
This work will address this question with the $T^+_{cc}$ and $\psi
(3770)$ ($^3D_1$) experimental data.

The $T_{cc}^{+}$ was recently reported by LHCb collaboration in the
$D^0 D^0 \pi^+$ channel~\cite{LHCb:2021vvq,LHCb:2021auc} and  has
attracted a great deal of interest~\cite{Agaev:2021vur,
Ling:2021bir, Chen:2021vhg, Dong:2021bvy, Feijoo:2021ppq,
Yan:2021wdl, Xin:2021wcr, Huang:2021urd, Fleming:2021wmk,
Azizi:2021aib, Hu:2021gdg, Chen:2021cfl, Albaladejo:2021vln,
Du:2021zzh, Deng:2021gnb, Agaev:2022ast, Braaten:2022elw,
He:2022rta, Abreu:2022lfy, Achasov:2022onn, Mikhasenko:2022rrl,
Wang:2022jop, Deng:2022cld, Lyu:2023xro}.
Since the $T^+_{cc}$ is located just below the $D^0 D^{*+}$
threshold within several hundred keV, the $DD^{*}$ interaction
definitely plays the most important role in the formation of this
exotic state, offering precious experimental data to constrain the
$DD^*$ interaction.

To combine the strength of experimental data with a direct
connection between $DD^*$ and $D\bar D^*$, we first perform a
detailed analysis of the $D^0D^0\pi$ invariant mass for the
$T^+_{cc}$ distributions considering the dynamical complexities of
the $DD^*$ interaction.
Our approach employs a coupled-channel formalism and incorporates
comprehensive scattering potentials, including the exchanges of
light mesons ($\pi$, $\rho$, $\omega$).
These $D\bar D^*$ potentials are derived from a heavy hadron
effective Lagrangian and $D\bar D^*$ interactions can be directly
obtained with the charge
conjugation symmetry.
The potential coupling of $c\bar{c}$ with $D\bar D^*$ happens
through the quark-pair creation (QPC) model, which is determined
through the well-studied $c\bar c$ state $\psi (3770)$ ($^3D_1$).

\section{Materials and methods}

We first concentrate on the $DD^*/ D\bar D^*$ interaction.
The S-wave $DD^*$ and $D\bar D^*$ may form the molecules with the
spin-partiy $J^{P(C)}=1^{+(\pm)}$ with $I=1, 0$, as listed in Table
\ref{tab:obe}\footnote{The wave functions presented in
Table I are shown to systematically describe all possible
resonances.  We did all the calculations on a physical particle
basis instead of the isospin eigenstates. Thus, our result is
independent of isospin eigenstates.}.
In the following, we simultaneously study the relatively long-range
$DD^*$/ $D\bar D^*$ interactions by exchanging the light mesons.
We focus on the light pseudoscalar (P) meson $\pi$- and vector (V)
meson $\rho/\omega$-exchange potentials for the $DD^*$/$D\bar D^*$,
while the $\sigma$- and $\eta$-contribution are neglected due to
their tiny contributions \cite{Li:2012ss,Li:2012cs,Wang:2018jlv}.
The interactions are derived using the Lagrangians given in
Refs.~\cite{Li:2012ss,Li:2012cs} based on heavy quark symmetry.
The $DD^*$ and $ D\bar D^*$ interactions are related to each other
using the charge conjugation.

\begin{table*}[t]
\renewcommand\arraystretch{1.8}
\caption{The $D^*D$ and $D\bar D^*$ interactions in the
one-boson-exchange model in the isospin symmetry limit. To
facilitate comparison, we have employed the potential forms
(${D^*D}$ interactions) to express both $D^*D$ and $D\bar D^*$
interactions, denoted as $V^{u/t}_{\phi}$ where the subscript $\phi$
may be $\pi$, $\rho$, $\omega$. The C-parity of the flavor wave
functions (neutral system) $[D\bar D^*]=\frac{1}{\sqrt{2}}(D\bar
D^*-D^*\bar D)$ and $\{D\bar D^*\}=\frac{1}{\sqrt{2}}(D\bar
D^*+D^*\bar D)$ are even and odd, respectively, using the charge
conjugation convention $D^*\rightarrow -\bar D^*$ from Ref.
\cite{Liu:2008fh}.
} \label{tab:obe}
\begin{ruledtabular}
\begin{tabular}{c|ccccc}
\multirow{1}{*}{} & wave function & $I(J^{PC})$ &
$u-\text{channel}:\,\ensuremath{\pi}$ &
$u-\text{channel}:\,\rho/\omega$ &
$t-\text{channel}:\,\rho/\omega$\tabularnewline \hline
\multirow{2}{*}{$DD^{*}$} & $\frac{1}{\sqrt{2}}(D^{+}D^{*0}-
D^{0}D^{*+})$ & $0(1^{+})$ [$T^+_{cc}$] & $\frac{3}{2}V_{\pi}$ &
$\frac{3}{2}V_{\rho}^{u}-\frac{1}{2}V_{\omega}^{u}$ &
$-\frac{3}{2}V_{\rho}^{t}+\frac{1}{2}V_{\omega}^{t}$\tabularnewline
& $\frac{1}{\sqrt{2}}(D^{+}D^{*0}+D^{0}D^{*+})$ & $1(1^{+})$ &
$\frac{1}{2}V_{\pi}$ &
$\frac{1}{2}V_{\rho}^{u}+\frac{1}{2}V_{\omega}^{u}$ &
$\ensuremath{\frac{1}{2}V_{\rho}^{t}+\frac{1}{2}V_{\omega}^{t}}$\tabularnewline
\hline \multirow{4}{*}{$D\text{\ensuremath{\bar D^{*}}}$} &
$\frac{1}{\sqrt{2}}\left(\left[D^{+}D^{*-}\right]+\left[D^{0}\bar{D}^{*0}\right]\right)$
& $0(1^{++})$[$X(3872)$] & $\frac{3}{2}V_{\pi}$ &
$-\frac{3}{2}V_{\rho}^{u}-\frac{1}{2}V_{\omega}^{u}$ &
$-\frac{3}{2}V_{\rho}^{t}-\frac{1}{2}V_{\omega}^{t}$\tabularnewline
&
$\frac{1}{\sqrt{2}}\left(\left[D^{+}D^{*-}\right]-\left[D^{0}\bar{D}^{*0}\right]\right)$
& $1(1^{++})$ & $-\frac{1}{2}V_{\pi}$ &
$\frac{1}{2}V_{\rho}^{u}-\frac{1}{2}V_{\omega}^{u}$ &
$\frac{1}{2}V_{\rho}^{t}-\frac{1}{2}V_{\omega}^{t}$\tabularnewline

& $ \frac{1}{\sqrt{2}}\left(\left\{ D^{+}D^{*-}\right\} +\left\{
D^{0}\bar{D}^{*0}\right\} \right)$ & $0(1^{+-})$[$h_c$] &
$-\frac{3}{2}V_{\pi}$ &
$\frac{3}{2}V_{\rho}^{u}+\frac{1}{2}V_{\omega}^{u}$ &
$-\frac{3}{2}V_{\rho}^{t}-\frac{1}{2}V_{\omega}^{t}$\tabularnewline

& $\frac{1}{\sqrt{2}}\left(\left\{ D^{+}D^{*-}\right\} -\left\{
D^{0}\bar{D}^{*0}\right\} \right)$ & $1(1^{+-})$ [$Z_c(3900)$] &
$\frac{1}{2}V_{\pi}$ &
$-\frac{1}{2}V_{\rho}^{u}+\frac{1}{2}V_{\omega}^{u}$ &
$\frac{1}{2}V_{\rho}^{t}-\frac{1}{2}V_{\omega}^{t}$\tabularnewline

\end{tabular}
\end{ruledtabular}
\end{table*}

The $DD^*$ and $D\bar{D}^*$ potentials by exchanging the light-meson
under the isospin bases are summarized in Table \ref{tab:obe}.
The potentials $V_{\pi}$, $V^{u}_{\rho/\omega}$ and
$V^{t}_{\rho/\omega}$ are given in terms of three coupling constant
$g$($D^*DP$), $\lambda$($D^*DV$), and $\beta$($DDV$/$D^*D^*V$),
respectively, as shown in the Appendix.
The constant $g=0.57$ is determined from the strong decay width of
$D^*\rightarrow D\pi$ \cite{ParticleDataGroup:2022pth}.

\begin{figure}[t]
\centering
\includegraphics[width=1.0\linewidth]{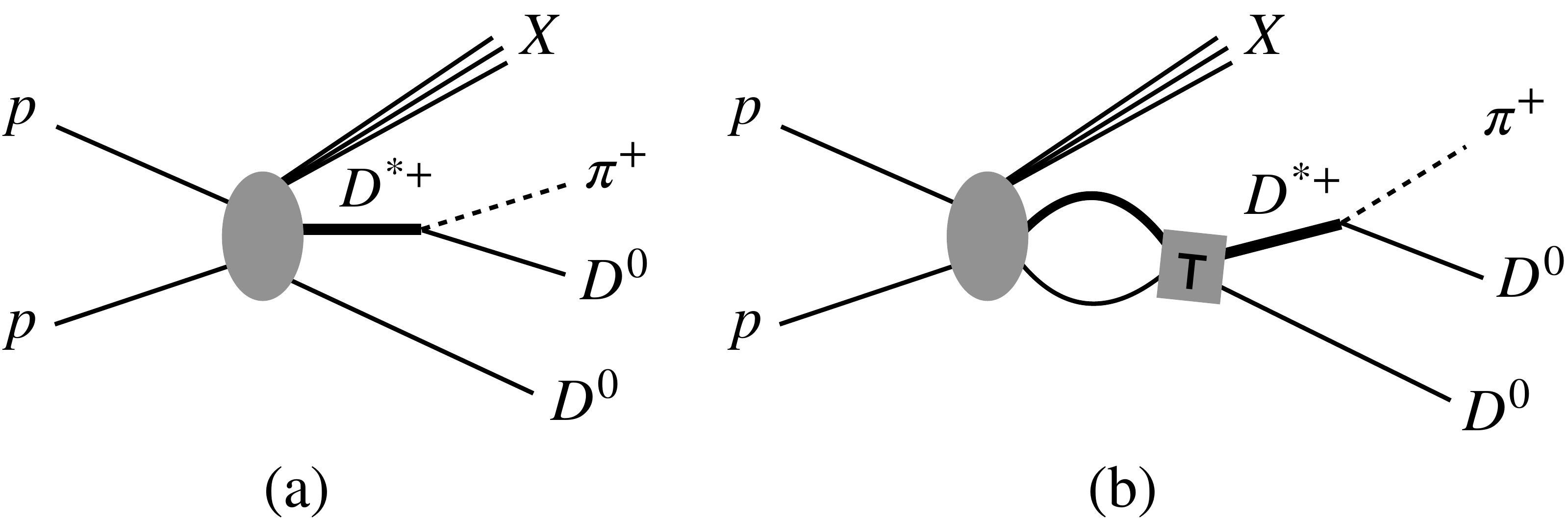}
\caption{The Feynman diagram for $pp\rightarrow D^0D^0\pi^+ X$($X$
denotes all the produced particles other than the $D^0$, $D^0$ and
$\pi^+$ in the collision). The square labeled as $T$ denotes the
scattering $T$-matrix for the $D^0D^{*+}$ and $D^+D^{*0}$ channels.}
\label{fig:tccpro}
\end{figure}

Subsequently, we determine the above $D D^{*} / D\bar{D}^* $
interaction by fitting the experimental $T^+_{cc}$ spectral function
with the contributions from the $D^0D^{*+}$ and $D^{*0}D^{+}$
channels.
The inclusive production of the $T^+_{cc}$ in the $pp\rightarrow
D^0(p_{D_1})D^0(p_{D_2})\pi^+ (p_\pi) X$, will be interpreted with
two mechanisms, as depicted in Fig.~\ref{fig:tccpro}.
The amplitude reads
\begin{widetext}
\begin{eqnarray} \label{eq:amplitude}
i\mathcal{M}_{pp\rightarrow DD\pi X} &=&\mathcal{A}^{\mu}_{pp\rightarrow DD^{*}X}\left\{ g_{\mu\alpha}-{\frac{i}{(2\pi)^{4}}}\int d^{4}q_{D^{*}}G_{D^{*}\,\mu\nu}(q_{D^{*}})G_{D}(p_{D_1}+p_{D_2}+p_\pi-q_{D^{*}})T^{\nu}_{\alpha}(q_{D^{*}},p_{D_1}+p_\pi)\right\} \nonumber \\
&&\times
G_{D^{*}}^{\alpha\beta}(p_{D_2}+p_\pi)(g\,p_{\pi,\beta})+(p_{D_1}
\rightarrow p_{D_2}),
\end{eqnarray}
\end{widetext}
where $\mathcal{A}_{pp\rightarrow DD^{*}X}^{\mu}$ is the production
vertex $pp\rightarrow DD^{*}X$.
In the energy region close to the threshold, we consider the $S$-wave $DD^*$ production contribution only. %
We have explored both the iso-vector and iso-scalar assignment for
the $\mathcal{A}$ with the production amplitudes satisfying
$\mathcal{A}_{pp\rightarrow
D^+D^{0*}X}^{\mu}=\pm\mathcal{A}_{pp\rightarrow D^0D^{*+}X}^{\mu}$,
respectively. We are able to find a satisfactory fit to the
experimental data only in the iso-scalar case.
The $G_H$ is the propagator of the heavy meson $H$.
The $DD^*$ scattering amplitude
$T(p_{D^*},p'_{D^*})\equiv\epsilon^*_\mu(p_{D^*}) T^{\mu
\nu}(p_{D^*},p'_{D^*})\epsilon_\nu(p'_{D^*})$ with $\epsilon$ the
polarization vector, can be solved from the relativistic
Lippmann-Schwinger
equation~\cite{Matsuyama:2006rp,Wu:2012md,Wu:2014vma,Liu:2015ktc},
\begin{eqnarray} \label{eq:lse}
&&T(\vec{p}_{D^{*}},\vec{p}^{\prime}_{D^{*}};E)={\mathcal
V}(\vec{p}_{D^{*}},\vec{p}^{\prime}_{D^{*}};E)+\int d\vec{q}
\nonumber\\
&&\,\,\times \frac{{\mathcal
V}(\vec{p}_{D^{*}},\vec{q};E)T(\vec{q},\vec{p}'_{D^{*}};E)}{E-\sqrt{m_D^2+q^2}-\sqrt{m_{D^*}^2+q^2}+i\epsilon}.
\end{eqnarray}
In our calculation, the $D^0D^{*+}$ and $D^+D^{*0}$ are two
independent channels. Thus the $T$ and $\cal V$-matrix  are both
$2\times2$ matrices.
The effective potential ${\mathcal V}$ is obtained with the
light-meson exchange potentials %  $V^{u/t}_{\phi}$
which incorporates a form factor to ensure regularization and
convergence,
\begin{equation}
{\mathcal
V}=\left(V_{\pi}+V^{t}_{\rho/\omega}+V^{u}_{\rho/\omega}\right)\left(\frac{\Lambda^{2}}{\Lambda^{2}+p_{f}^{2}}\frac{\Lambda^{2}}{\Lambda^{2}+p_{i}^{2}}\right)^{2},\label{eq:ff}
\end{equation}
where $\Lambda$ is the cut-off parameter.
The introduction of the form factor can suppress the
$\delta$ potential from the one-pion-exchange.
\section{Results}

\begin{figure}[t]
\centering
\includegraphics[width=0.9\linewidth]{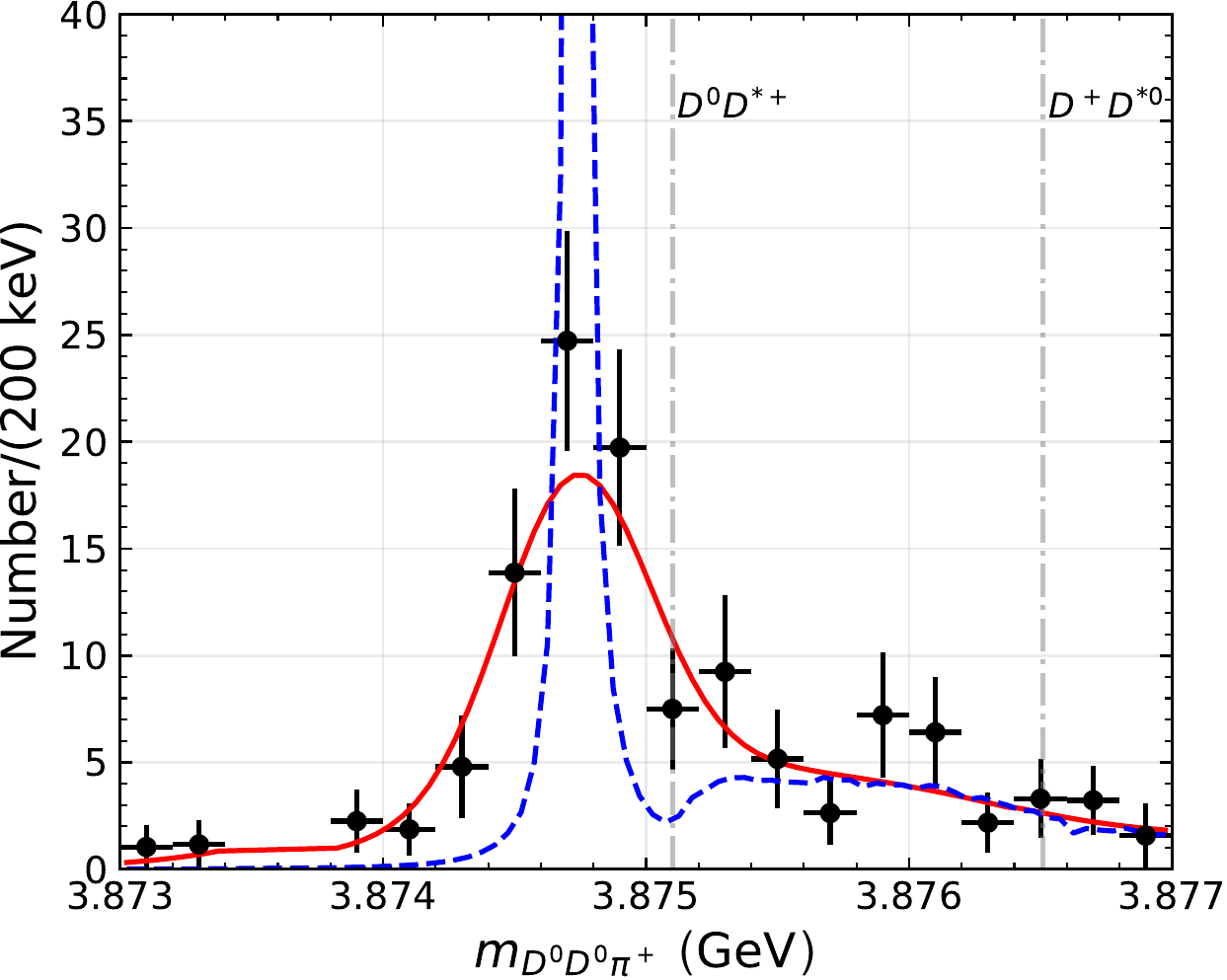}

\caption{ (color online) The fitted lineshape of the $T^+_{cc}$ in the
$D^0D^0\pi^+$ invariant mass spectrum~\cite{LHCb:2021vvq} .  The
blue dashed and red solid lines represent the lineshapes before and
after the convolution with the energy resolution function, which is
taken from the LHCb collaboration~\cite{LHCb:2021auc}. }
\label{fig:fitexp}
\end{figure}

The cutoff reflects the composite structure of related
hadrons and the off-shell effect.
However, we expect the results to be insensitive to the specifics of
hadronic structure or off-shell effects. We should check the cutoff
dependence in a reasonable range, around 0.8 GeV to 1.2 GeV.
Only adjusting the cutoff value impacts the $T$-matrix value due to
its role in compensating part of the UV divergence contribution.
Thus, we expect that the cutoff dependence should be effectively
absorbed into the coupling constants $\lambda$ and $\beta$ by
fitting the data.
To confirm this, we perform the fitting procedures with three values
$\Lambda=0.8$, $1.0$, and $1.2$ GeV, respectively,
 to check the quality of the three fits and the uncertainty of the attracted $T_{cc}$ pole position.
Indeed, our final results remain unchanged regardless of the chosen
$\Lambda$ value.
When the cutoff $\Lambda=1.0$ GeV is taken, the fitted parameters
are
\begin{equation} \label{eq:para}
\lambda=0.683\pm0.025\,\text{/GeV}, \quad\beta=0.687\pm0.017.
\end{equation}
with $\chi^2/\text{d.o.f.}=0.78$.
The fitted line shape is shown in Fig.~\ref{fig:fitexp}, together
with the line shape before the convolution with the energy
resolution function~\cite{LHCb:2021auc}.

\begin{figure}[t]
\centering
\includegraphics[width=0.9\linewidth]{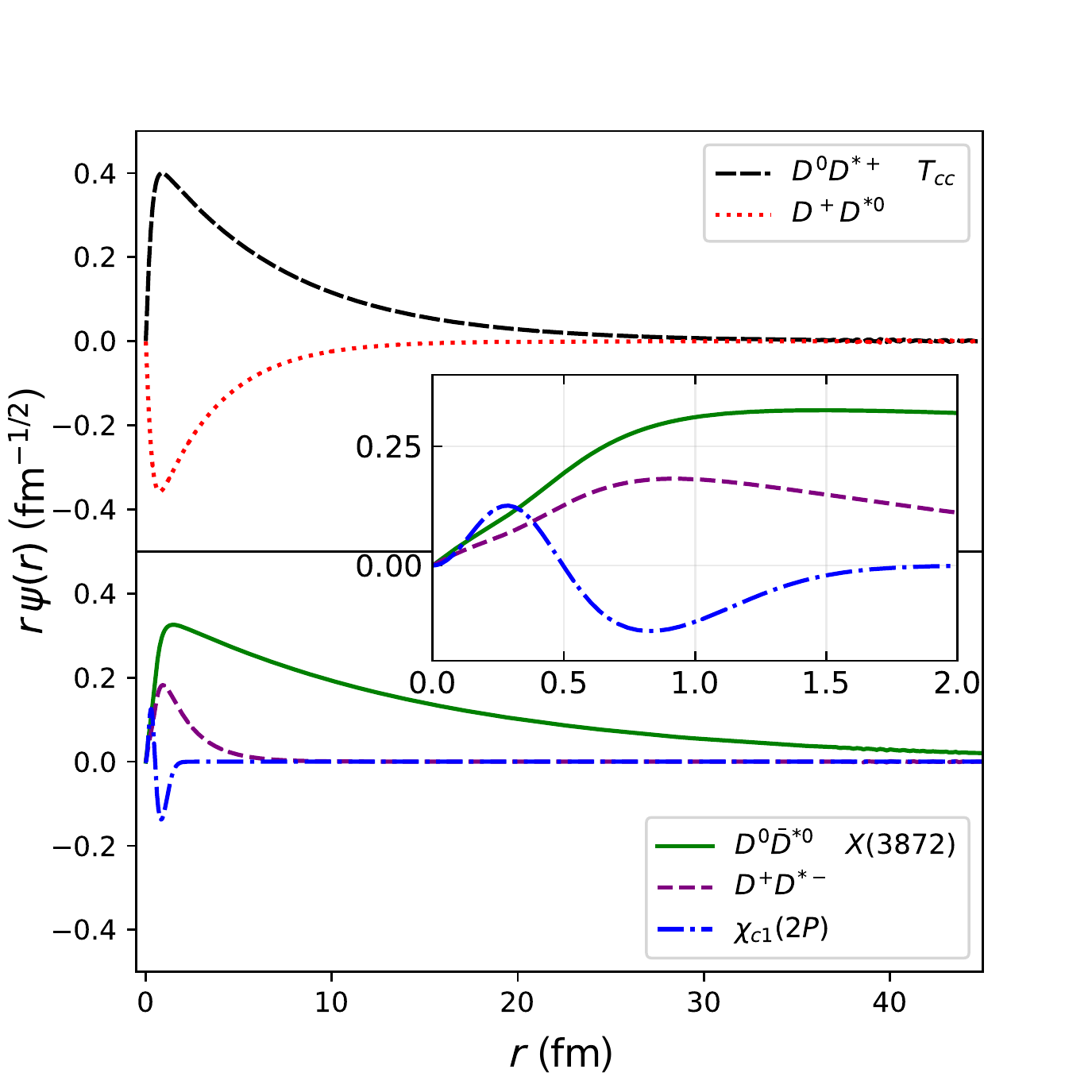}

\caption{ (color online) The wave function distribution of different components for
the $T^+_{cc}$ and $X(3872)$. } \label{fig:wavefun}
\end{figure}

To identify $T^+_{cc}$ and explore other possible resonant states
while obtaining the wave function, we search the poles in the
complex plane using $T$-matrix pole analysis and complex scaling
method (CSM)~\cite{Aoyama:2006,Myo:2014ypa,moiseyev1998quantum},
which is introduced in the Appendix. The results are consistent with
each other on the bound and resonant states. Our results clearly
show a distinct signal corresponding to the bound state of the
$DD^{*}$. We summarize its properties including the mass, decay
width, root mean square radius and proportions of different
components in Table \ref{tab:Tcc}.
The binding energy of the bound state is $\Delta E=-393.0$ keV,
which is consistent with that of the experiment $\Delta
E_{\text{exp}}=-360(40)$ keV~\cite{LHCb:2021auc}. In our model, the
whole calculation is conducted in the momentum space and the
on-shell contribution is included, which naturally generates the
width of the $T^+_{cc}$ around $70$ keV through the decay into the
$DD\pi$ final state. In Ref.~\cite{Du:2021zzh}, the $D^*$ width in the intermediate loops was found to have nonnegligible contribution to the $T^+_{cc}$ width. In our present calculation, the width of  $D^*$ is not included since our basic model focuses on the two-body system, rather than $\pi D D$ system.  The impact of the three-body effects will be studies in the future under the same framework.

The wave function of $T^+_{cc}$ is presented in Fig.
\ref{fig:wavefun}.
As a shallow S-wave bound state, there is a long tail for the radius
distribution.
The root mean square radius is around $4.7$ fm which establishes the
$T^+_{cc}$ as a molecular state of the $DD^{*}$.
The ratio of the residue of the $D^{*+}D^0$ and $D^{+}D^{*0}$
channels is close to $1$, showing the similar couplings of the
$T^+_{cc}$ to these two components.
Then the difference between the wave functions of two channels is
mainly due to their mass difference of $1.4$ MeV.
Such small mass splitting still introduces a sizeable isospin
breaking effect because of its extremely small binding energy, and
the iso-vector component occupies around $4.2\%$ in $T^+_{cc}$ as
shown in Table \ref{tab:Tcc}.

\begin{table*}
\centering \caption{ The properties of the $T^+_{cc}$ and $X(3872)$
in the fit with $\Lambda=1.0$ GeV. The script ``BE" denotes the
binding energy. The ratio of the residue in two channels is listed
in the last column. For $X(3872)$, the QPC parameter $\gamma =
4.69$. } \label{tab:Tcc}
\begin{ruledtabular}
\begin{tabular}{c|ccccccccc}

 & BE (keV) & $\Gamma$ (keV) & $\sqrt{\langle r^2\rangle}$ & $I=0$ & $I=1$ & $P(D^0D^{*+})$ & $P(D^{+}D^{*0})$&  $|\frac{\text{Res}(D^0D^{*+})}{\text{Res}(D^{+}D^{*0})}|$\\
 \hline
$T^+_{cc}$ & -$393.0$ &$70.4$ &  $4.7$ fm & $95.8\%$ & $4.2\%$ & $70.0\%$  &$30.0\%$  & $1.055 $\\
 \hline  \hline
& BE (keV) & $\Gamma$ (keV) & $\sqrt{\langle r^2\rangle}$ & $I=0$ & $I=1$ & $P(D^0\bar D^{*0})$ & $P(D^{+} D^{*-})$ & $P(c\bar{c})$ %&  $\frac{\text{Res}(D^0\bar D^{*0})}{\text{Res}(D^{+}D^{*-})}$\\
\\ \hline
%$0.28\sqrt {96\pi}$& -172.7 & 41.0 & 7.8 & 76.1\% & 23.9\% & 91.1\% & 7.0\% & 1.9\% \\%&
$X(3872)$  & -80.4 & 32.5 & 11.2 fm & 71.9\% & 28.1\% & 94.0\% & 4.8\% & 1.2\% %&

\end{tabular}
\end{ruledtabular}
\end{table*}

The successful interpretation of the $T^+_{cc}$ encourages us to
study the molecular states in the $D\bar D^*$ sector composed of the
$\bar D^0D^{*0}$ and $D^+D^{*-}$.
With the $DD^*$ interactions, we derive the $D\bar D^*$ interactions
through the charge conjugation.
Then in the $J^{PC}=1^{++}$ sector where the $X(3872)$ exists, we do
not find any bound state, but we
obtain a virtual state with $E_v= 3870.0 +0.26i$ MeV in the second
Riemann sheet of the neutral channel and the first Riemann sheet of
the charged channel.
Interestingly, if we introduce a scaling factor ($>1$), for example
$1.15$, to all couplings of the $D^*\bar D$ potentials, a bound
state appears with a binding energy $-25$ keV.
This indicates that the $D\bar D^*$ interaction is attractive but
not strong enough to produce a bound state. The conclusion remains
the same using the different cut off value such as $0.8$ or $1.2$
GeV.
Moreover, the bound state can be obtained by fine-tuning the
coupling strength in the OBE model as shown above.
Therefore the constraint from the $T^+_{cc}$ is indispensable to
study the formation mechanism of the $X(3872)$.

In fact, the $D\bar D^*$ system with the quantum numbers
$I(J^{PC})=0(1^{++})$  can couple with the $\chi_{c1}(2P)$.
In the following, we demonstrate that the inclusion of the bare
$c\bar c$ core is essential to form the $X(3872)$.
The mass and wave functions of  $\chi_{c1}(2P)$ are determined by
using the GI model with the updated parameters from our previous
works~\cite{Yang:2021tvc,Yang:2022vdb}.
To account for the coupled channel effect between the $c\bar c$ core
and $D \bar D^{(*)}$, we employ the quark-pair-creation (QPC) model.
Within the framework, a light quark pair with the spin-parity
{$J^{PC}=0^{++}$} is created and combined with the $c\bar c$ pair to
form the $D\bar D^*$ state.
This approach provides a feasible connection between the quark and
hadron level.
The transition potential reads
\cite{LeYaouanc:1977fsz,Kokoski:1985is,Page:1995rh,Blundell:1996as,Ackleh:1996yt,Morel:2002vk,
Ortega:2016mms},
 \begin{equation}
g_{D\bar D^*,c\bar c }(|\vec{k}_{D\bar D^*}|)=\gamma I_{D\bar D^*,c\bar c }(|\vec{k}_{D\bar D^*}|), %e^{-\frac{\vec{k}^2_{D\bar D^*}}{2\Lambda^{\prime 2}}},
\end{equation}
where $\vec{k}_{D\bar D^*}$ is the relative momentum in the ${D\bar
D^*}$ channel.
$I_{D\bar D^*,c\bar c }(|\vec{k}_{D\bar D^*}|)$ is the overlap of
the meson wave functions.
The parameter $\gamma$ represents the amplitude of producing the
light quark pair.
We choose $\gamma=4.69$ so that it reproduces the mass and decay of
$\psi(3770) (\to D\bar D)$ resonance.
There may be additional contributions for the $D\bar D^*$
interactions, such as the exchange of heavy mesons and their
excitations, leading to the coupling $D\bar D^*-\text{charmonium}
+\text{light\, mesons}$.
In the molecular scenario, such contributions to the $D\bar D^*$
rescattering effects are significantly suppressed compared with
those originating from the  $D\bar D^*-c\bar c$ coupling and not
included in this work.

Using the CSM method, our results show a clear signal of a bound
state associated with the $X(3872)$ with a binding energy $\Delta E
= -80.4$ keV and a resonant state corresponding to a dressed $c\bar
c$ state with $J^{PC}=1^{++}$.
%   radius of approximate $7.8$ fm, and

In Table \ref{tab:Tcc}, we summarize the properties of $X(3872)$.
Its width, $32.5$ keV,  arises predominantly from the decay into the
$D^0\bar{D}^0\pi^0$ channel.
The dominant component of the $X(3872)$ is the loosely bound
molecular state of the neutral $D^0\bar{D}^{*0}$ ($94.0\%$) and
charged $D^+\bar{D}^{*-}$~($4.8\%$).
The mass difference of $8$ MeV between the charged and neutral
channels leads to an important isospin breaking for a bound state
extremely close to the neutral channel.  It is worth highlighting
that the $c\bar{c}$ component is crucial to form the $X(3872)$, even
though it only accounts for $1.2\%$.

To visualize the contribution of the $c\bar{c}$, $D^+D^{*-}$ and
$D^0\bar{D}^{*0}$ components in the $X(3872)$, we display their wave
functions in the lower and middle panels of Fig. \ref{fig:wavefun}.
We employ the $c\bar{c}$ wave function from the GI model, combing
with its probability amplitude in the $X(3872)$.
%multiplied by the relative wave function coefficient in the
%$X(3872)$.
In the short-range region up to $2$ fm (middle panel of Fig.
\ref{fig:wavefun}), both the $c\bar c$ and $D\bar D^{*}$ components
are significant.
Notably, for $r <0.5$ fm, the $c\bar c$ core dominates.
The wave functions show that the main component $D\bar D^{*}$
definitely plays the dominant role in the long-distance region,
which contributes to the large size of the $X(3872)$ whose radius
around $11.2$ fm. This is reasonable because the $X(3872)$ has such
a tiny binding energy and couples strongly with the $D\bar D^{*}$
channel.

The isospin-breaking decays $X(3872)\rightarrow J/\psi \omega/ \rho$
is attributed to the overlap factor of the wave functions $\int d
\vec{p} \psi^{I=1/0}_{X(3872)}(\vec{p})$ $\psi^*_{J/\psi}(\vec{p})$
in the amplitude \cite{Braaten:2003he}, yielding a ratio
$R_{\omega/\rho}=\frac{\mathcal{B}[X\rightarrow J/\psi
\omega]}{\mathcal{B}[X\rightarrow J/\psi \rho]}=21.4$. Using the
factorization formulae
\cite{Braaten:2005ai,Gamermann:2009uq,Meng:2021kmi}, the decay ratio
$\frac{\mathcal{B}[X\rightarrow J/\psi
\pi^+\pi^-\pi^0]}{\mathcal{B}[X\rightarrow J/\psi
\pi^+\pi^-]}=R_{\omega/\rho}\times R_2$ is $1.86$  and $3.14$, with
$R_2\sim \frac{\mathcal{B}(\omega \rightarrow
\pi^+\pi^-\pi^0)}{\mathcal{B}(\rho\rightarrow \pi^+\pi^-)}$ being
$0.087$ \cite{Braaten:2005ai} and $0.147$
\cite{Gamermann:2009uq,Meng:2021kmi}, respectively. The former decay
ratio is consistent with the  experimental value $1.0\pm 0.4\pm0.3$
\cite{Belle:2005lfc}.
Besides the $X(3872)$, we also find a signal of the resonant state
$\chi_{c1}(2P)$ at
\begin{equation}
    M=3957.9 \, \text{MeV},\quad  \Gamma=16.7 \, \text{MeV},
\end{equation}
which might  be identified as the $X(3940)$($M=3942\pm 9$ MeV and
$\Gamma=37^{+27}_{-17}$ MeV) observed in the $D\bar D^*$ channel
\cite{Belle:2005lik}.
Moreover, in the $J^{PC}=1^{+-}$ sector,  we investigate the
coupling between the $h_c(2P)$ component and iso-scalar $D\bar D^*$
channel. Two poles are observed. One is a virtual state around
$M=3870.2$ MeV. Since the $t$-channel vector meson exchange
potentials dominate, it is related to an isoscalar state $\tilde
X(3872)$ with a mass $3860.0\pm10.4$ MeV discovered by COMPASS
Collaboration~\cite{COMPASS:2017wql}. The existence of $\tilde
X(3872)$ is also post-predicted by several theoretical
investigations~\cite{Wang:2020dgr,Dong:2021juy,Ortega:2021yis}.
Another signal corresponds to the excited $h_c(2P)$ whose pole is
around $M=3961.3$ MeV with the width $\Gamma=1.1$ MeV.

\section{Discussion and conclusion}

In this work, we first analyze the $DD\pi$ invariant mass
distributions for the $T^+_{cc}$ in the molecular scenario using a
coupled-channel formalism. We derived the scattering potential
involving the light mesons ($\pi$, $\rho$, $\omega$) from a heavy
hadron effective Lagrangian.
The data can be well described.
The one-boson-exchange $DD^*$ potentials enable a direct connection
to the $D\bar D^*$ potentials via the charge conjugation symmetry.
Thus, we set up a universal framework for the $DD^*$/ $D\bar D^*$
molecules.
It is different from the other frameworks which have the contact
constant potentials.
Although they can well fit the experimental data, all the dynamical
information is encoded in the parameters.
Notably, the short-range interactions in the $D\bar D^*$ and $D D^*$
are distinct due to the potential annihilation of the light and
heavy quark-antiquark pair in the $D\bar D^*$ system.
As a result, the direct application of charge conjugation symmetry
to derive the contact potentials for the $D\bar D^*$ from the $D
D^*$ sector is not feasible.
In our framework, the differences of the short-range
interaction of $D\bar{D}^*$ and $DD^*$ mainly arises from the
significant role of the charmonium state which couples with the
$D\bar{D}^*$.
This assumption promises that the OBE model can be fairly applied in
both sectors through the charge conjugation symmetry.

Within our framework, we explore two fine-tuning mechanisms of the
$X(3872)$.
Our results show that the pure $D\bar D^*$ interaction is attractive
but insufficient to form a bound state $X(3872)$.
The P-wave $c\bar{c}$ core plays a crucial role in the formation of
the $X(3872)$, even though its proportion is quite small.
Notably, the wave function of the $X(3872)$ shows that the $D \bar
D^*$ component has a long tail causing its large radius, while the
$c\bar c$ is significant at short distances.
Furthermore, our study successfully identifies three clear signals
related to the $\tilde X(3872)$, $X(3940)$, and  $h_c(2P)$ states.
The $X(3940)$ which was first observed in $D\bar D^*$ channel awaits
further exploration, and the $h_c(2P) $ state remains to be
discovered. These findings can serve as benchmarks for evaluating
our new model in future experiments.

Notably, this framework can be extended to establish other exotic
states, for which no consensus has been reached for their dynamic
origins.
They often exhibit characteristics that come from the mixing of
various dynamic sources.
Therefore, it is important and useful to identify and study their
dynamics from their relations to the well-understood hadronic
states.
Such a systematic approach may allow us to unravel the intricate
structure of the exotic hadronic states, providing a unique and
reliable pathway for investigating the $XYZ$ particles.

\begin{acknowledgments}

We thank useful discussions and valuable comments from Dote Akinobu,
Meng-Lin Du, Jian-Bo Cheng, Rui Chen, Feng-Kun Guo, Zi-Yang Lin, Lu
Meng, Mao-Jun Yan, Bing-Song Zou. This work is partly supported  by
the KAKENHI under Grant Nos.~19H05159, 20K03959, and 21H00132
(M.O.), and 23K03427 (M.O. and G.J.W), and by the National Natural
Science Foundation of China (NSFC) under Grants Nos.~12275046
(Z.Y.), 12175239 and 12221005 (J.J.W), 11975033 and 12070131001
(S.L.Z.), and by the Natural Science Foundation of Sichuan Province
under Grant No. 2022NSFSC1795 (Z.Y.), and by the National Key R$\&$D
Program of China under Contract No. 2020YFA0406400 (J.J.W), and by
the Chinese Academy of Sciences under Grant No. YSBR-101 (J.J.W).

\end{acknowledgments}

\bigskip

Author contributions:\\
Jia-Jun Wu proposed the main idea in this project. Guang-Juan Wang and Zhi Yang did the calculations and drafted the manuscript. Makoto Oka and Shi-Lin Zhu refined the idea and checked the results. All the authors made substantial contributions to the discussions and the editing of the manuscript. All the authors have approved the final version of this manuscript.

\bigskip

\appendix

\section{Fit details}
The differential cross section for the $pp\rightarrow D
(p_{D_1})D(p_{D_2})\pi(p_\pi)$ channel reads
\begin{eqnarray} \label{eq:diff}
 d\sigma_{pp\rightarrow XDD\pi}    &=& \frac{(2\pi)^{4}}{4\sqrt{(p_{p_{1}}\cdot p_{p_{2}}-m_{p}^{2}m_{p}^{2})}}|\mathcal{M}|^{2}d\Phi_{XDD\pi} \nonumber \\
\frac{d\sigma_{pp\rightarrow XDD\pi}}{dm_{DD\pi}} &\thickapprox& 2m_{DD\pi}\int d\sigma_{pp\rightarrow X+DD\pi}B_{2}d\Phi_{DD\pi} \nonumber \\
&\thickapprox& 2m_{DD\pi} \int dm_{12}dm_{23}
B_{2}(E;m_{12},m_{23}), \nonumber \\
\end{eqnarray}
where $p_{p_1}$ ($p_{p_2}$) and $m_p$ are the momentum and mass of
the initial photon. $m_{ij}$ is defined as $m_{ij}^2=(p_i+p_j)^2$
with $p_i$ the momutum of $D$ or $\pi$ in the final state. $ B_{2}$
is obtained with the $\mathcal{M}$,
\begin{eqnarray}
|\mathcal{M}|^{2}&=&|a_{pp\rightarrow DD^{*}X}|^{2}B_2,\nonumber \\
B_2&=&\sum_{\lambda_X}\epsilon_{\mu}(p_X,\lambda_X)\epsilon^{\dagger}_{\mu'}(p_X,\lambda_X)\mathcal{B}{\mu}\mathcal{B}^{\dagger\mu'}.
\end{eqnarray}
To obtain Eq. \eqref{eq:diff} in the energy region close to the
threshold, we have approximated $\mathcal{A}_{pp\rightarrow
DD^{*}X}^{\mu}=a_{pp\rightarrow
DD^{*}X}\epsilon^{\mu}(p_X,\lambda_X)$ with $a_{pp\rightarrow
DD^{*}X}$ a constant and independent of the polarization. Moreover,
the scattering cross section $\sigma_{pp\rightarrow X+DD\pi}$ is a
constant since we only concentrate on a small energy range. For a
specific channel, the $\mathcal{B}^{\mu}_j$ (the $j$-th represents
the $D^*D$ channels) reads
\begin{widetext}
\begin{eqnarray}
\mathcal{B}_j^{\mu}(p_{12},p_{23})&=&g\left\{\frac{-i(p_{\pi}^{\mu}-\frac{p_{12}^{\mu}p_{12}\cdot
p_{\pi}}{m_{D^{*}}^{2}})}{p_{12}^{2}-m_{D^{*}}^{2}+im_{D^{*}}\Gamma_{D^*}}\right\}_j
+\sum_{i=1,2}ig\int dq_{D^{*}}q_{D^{*}}^{2}\frac{d\Omega_{q_{D^{*}}}}{4\text{\ensuremath{\pi}}}\frac{\sqrt{2w_{D_{2}}}}{\sqrt{2w_{D^{*}}}}\frac{\sqrt{2w_{D_{12}^{*}}}}{\sqrt{2w_{D}}}\nonumber\\
&\times&\frac{T_{ij}^{J00}(M,|q_{D^{*}}|,|p_{12}|)}{(M-w_{D^{*}}^{i})-w_{D}^{i}+i\epsilon}\left\{
\frac{\epsilon_{a}^{^{*}\mu}(w_{D^{*}},q_{D^{*}})\epsilon_{a}(p_{12})\cdot
p_{\pi}}{p_{12}^{2}-m_{D^{*}}^{2}+im_{D^{*}}\Gamma_{D^*}}\right\}_j
+(p_{D_1} \rightarrow p_{D_2}).
\end{eqnarray}
\end{widetext}
where the notations are $p_{12}=p_{D_1}+p_\pi$,
$p_{23}=p_{D_2}+p_\pi$ and $w_{H}=m_{H}^{2}+p_{H}^{2}$. In the
$D\bar D^*$  scattering, we only consider the S-wave contribution.
The $T_{i1}^{J00}$ is derived after the partial
wave decomposition of the $T(\vec{k}_{D^{*}},\vec{k}^{\prime}_{D^{*}};E)$. %To construct the $T$-matrix, we derive the potential from the interacting Lagrangian~\cite{Li:2012ss,Li:2012cs},

The explicit forms of the potentials in constructing the $T$-matrix
read
\begin{eqnarray}
V_{\pi}&=&\frac{g^{2}}{f_{\pi}^{2}}\frac{\left(q\cdot\epsilon_{\lambda}\right)\left(q\cdot\epsilon_{\lambda'}^{\dagger}\right)}{q^{2}-m_{\pi}^{2}},\\
{V}_{\rho/\omega}^u&=&-2\lambda^{2}g_V^{2}\frac{(\epsilon_{\lambda'}^{\dagger}\cdot q)(\epsilon_{\lambda}\cdot q)-q^{2}\left(\epsilon_{\lambda}\cdot\epsilon_{\lambda'}^{\dagger}\right)}{q^{2}-m_{\rho/\omega}^{2}},\\
V_{\rho/\omega}^t&=&\frac{\beta^{2}g_V^{2}}{2}\frac{\left(\epsilon_{\lambda}\cdot\epsilon_{\lambda'}^{\dagger}\right)}{q^{2}-m_{\rho/\omega}^{2}}.
\end{eqnarray}
Here we introduce a constant $g_V=5.8$ to compare our parameters
with other values from the phenomenological estimation
\cite{Li:2012ss,Li:2012cs,Wang:2018jlv}.

\begin{table*}[t]
\centering \caption{ The properties of the
$T^+_{cc}$~\cite{LHCb:2021vvq} in the three fits with the cutoff
values $\Lambda=0.8$, $1.0$, and $1.2$ GeV. The script ``BE" denotes
the binding energy. The ratio of the residue in two channels is
listed in the last column.} \label{tab:allTcc}
\begin{ruledtabular}
\begin{tabular}{c|ccccccccc}

$\Lambda$ (GeV) & BE (keV) & $\Gamma$ (keV) & $\sqrt{\langle
r^2\rangle}$ & $I=0$ & $I=1$ & $P(D^0D^{*+})$ & $P(D^{+}D^{*0})$&
$\frac{\text{Res}(D^0D^{*+})}{\text{Res}(D^{+}D^{*0})}$\\ \hline
$0.8$ &  -$387.7$  &$67.3$ &  $4.8$ fm & $95.8\%$ & $4.2\%$ & $70.0\%$  &$30.0\%$  & $-1.063+0.001I$ \\
$1.0$ & -$393.0$ &$70.4$ &  $4.7$ fm & $95.8\%$ & $4.2\%$ & $70.0\%$  &$30.0\%$  & $-1.055+0.001I $\\
$1.2$ & -$391.6$ & $72.7$ &  $4.7$ fm & $95.7\%$ & $4.3\%$ & $70.3\%$  &$29.7\%$  & $-1.052+0.001I $\\
%\hline\hline
%$\Lambda$ (GeV) & BE (keV) & $\Gamma$ (keV) & $\sqrt{\langle r^2\rangle}$ & $I=0$ & $I=1$ & $P(D^0D^{*0})$ & $P(D^{+}D^{-0})$ & $P(c\bar{c})$ %&  $\frac{\text{Res}(D^0D^{*0})}{\text{Res}(D^{+}D^{*-})}$\\
\end{tabular}
\end{ruledtabular}
\end{table*}

\section{The Independence of the cut-off $\Lambda$}
\begin{figure*}[t]
\begin{center}
\subfigure[]{
\begin{minipage}{0.5\linewidth}
\centering
\includegraphics[width=0.8\textwidth]{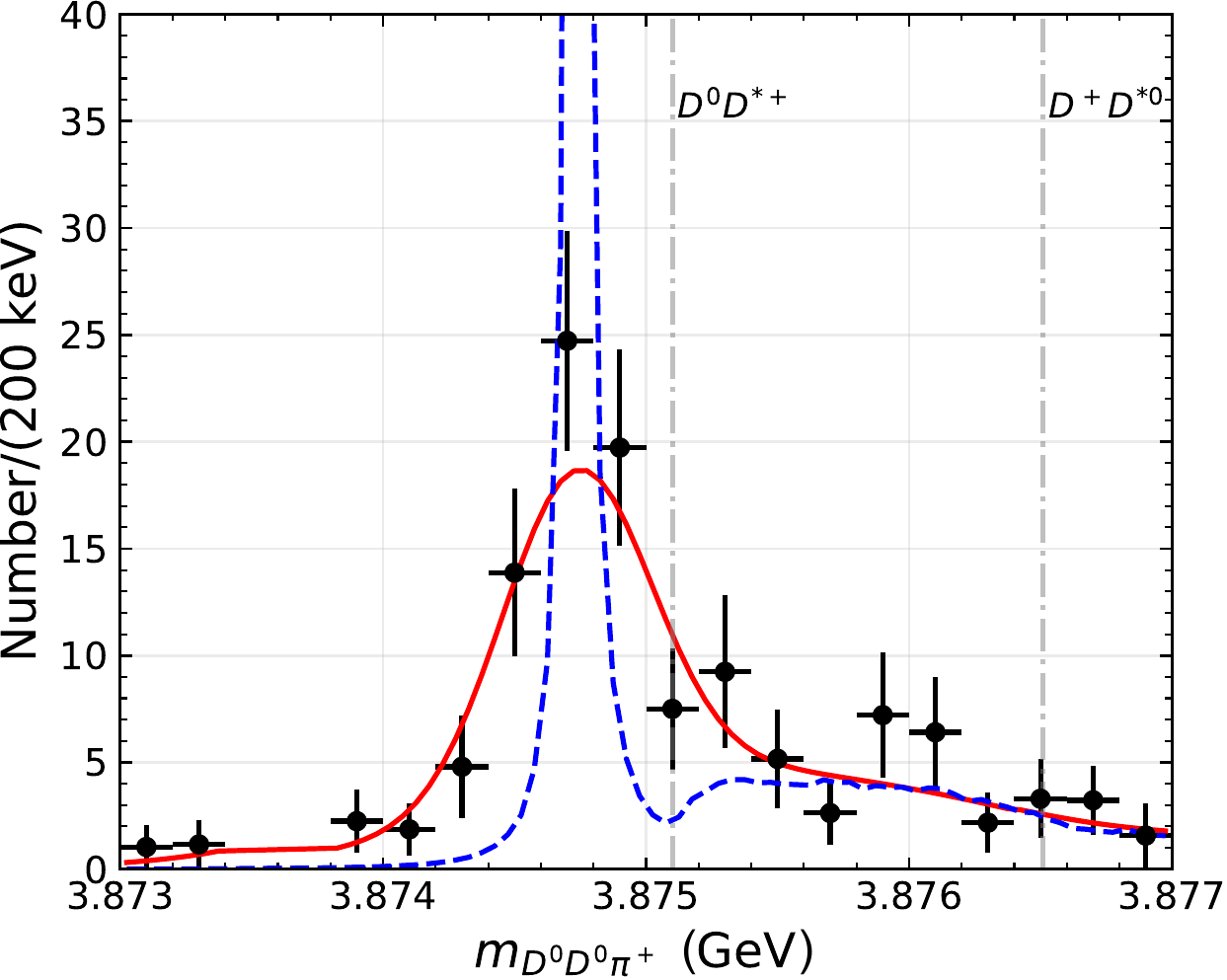}
\end{minipage}%
}%
%\subfigure[]{
%\begin{minipage}[t]{0.3\linewidth}
%\centering
%\includegraphics[width=\textwidth]{fit10.pdf}
%\end{minipage}%
%}%
\subfigure[]{
\begin{minipage}{0.5\linewidth}
\centering
\includegraphics[width=0.8\textwidth]{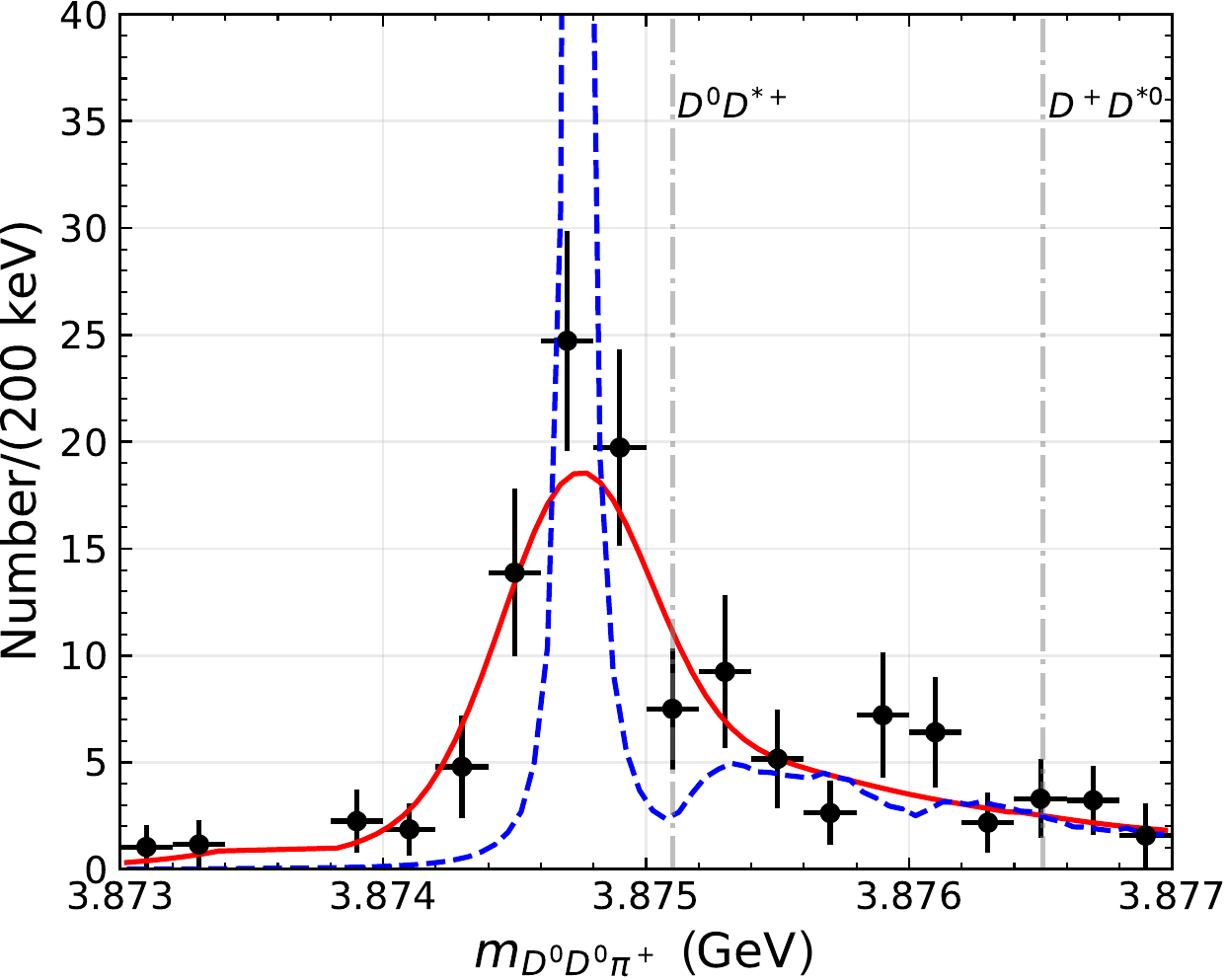}
\end{minipage}%
} \caption{(color online) The fitting results by employing the
light-meson-exchanging potential with the cutoff (a) $\Lambda=0.8$
GeV, (b)$1.2$ GeV. The $\chi^2/\text{d.o.f.}$ are  $0.76$ and
$0.78$, respectively. } \label{fig:fitexp2}
\end{center}
\end{figure*}

Besides the values $\Lambda=1.0$ GeV, we also used other two
different values $0.8$ and $1.2$ GeV for $\Lambda$. The lineshapes
fitted with $\Lambda=0.8$ and $1.2$ GeV are shown in
Fig.~\ref{fig:fitexp2}. As one can see the obtained results are
almost the same. The fitting parameters are
\begin{eqnarray}  \label{eq:para08}
&&
%\lambda_{\Lambda=0.8}=0.890\pm0.200/\text{GeV},\quad\beta_{\Lambda=0.8}=0.810\pm0.110. \nonumber\\
\lambda_{\Lambda=0.8}=0.890/\text{GeV},\quad\beta_{\Lambda=0.8}=0.810; \nonumber\\
&&
%\lambda_{\Lambda=1.2}=0.587\pm0.027/\text{GeV},\quad\beta_{\Lambda=1.2}=0.550\pm0.027.
\lambda_{\Lambda=1.2}=0.587/\text{GeV},\quad\beta_{\Lambda=1.2}=0.550.
\end{eqnarray}
The properties of the $T^+_{cc}$  with the fitting parameters are
summarized in Table \ref{tab:allTcc}, which clearly shows the
independence of the results on the values of the cutoff $\Lambda$.

\section{Complex scaling method}

\begin{figure*}[t]
\begin{center}
\subfigure[]{
\begin{minipage}{0.5\linewidth}
\centering
\includegraphics[width=0.8\textwidth]{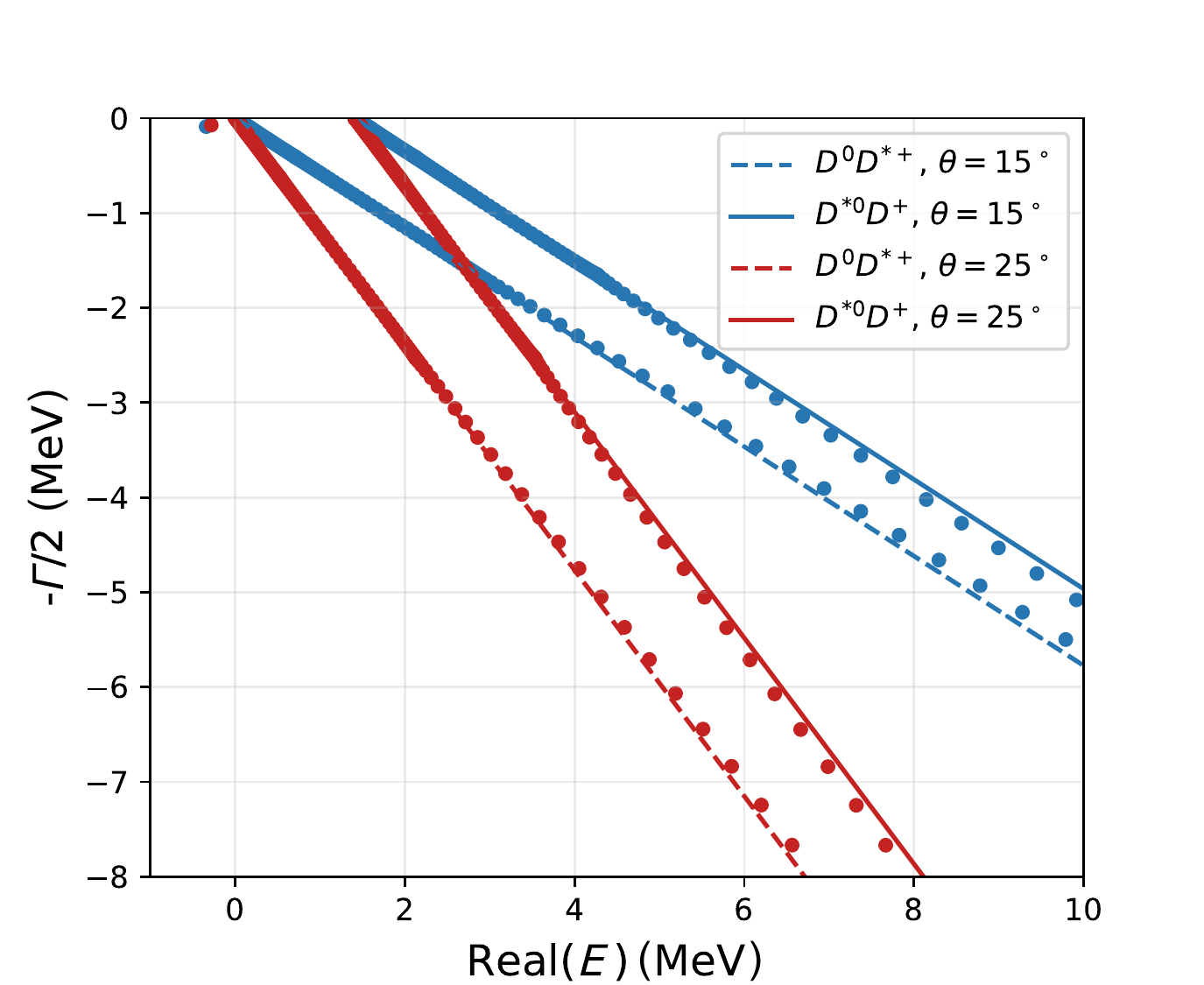}
\end{minipage}%
}%
\subfigure[]{
\begin{minipage}{0.5\linewidth}
\centering
\includegraphics[width=0.8\textwidth]{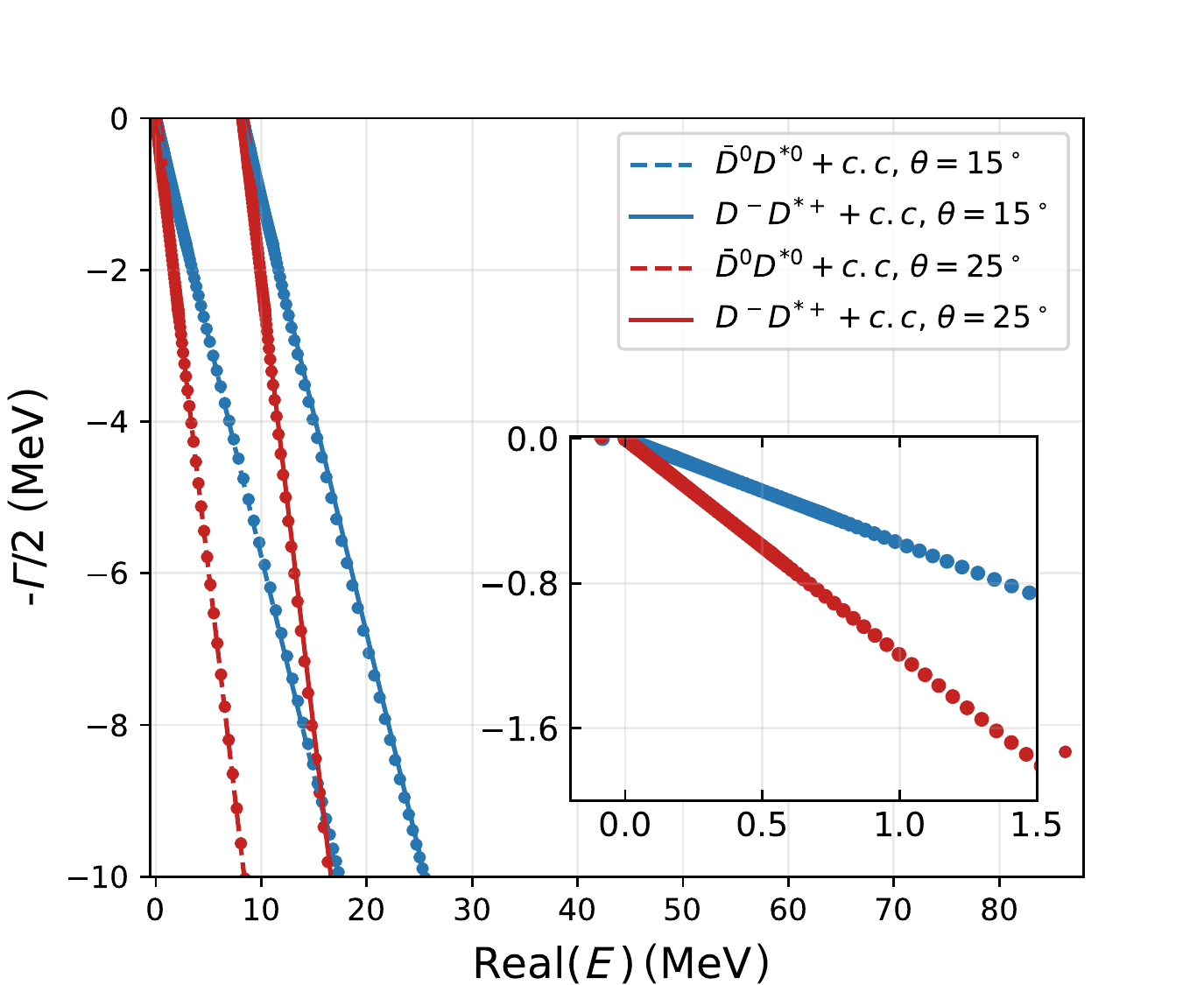}
\end{minipage}%
} \caption{(color online) The complex eigenvalues by employing the
light-meson-exchanging potential with the cutoff $\Lambda=1.0$ GeV
in the channels: (a) $D^0D^{*+}$ and $D^{*0}D^+$ channels; (b)
$\chi_{c1}(2P)$, $\bar D^0D^{*0}$ and $D^{-}D^{*+}$ channels. The
bound states in (a) and (b) correspond to the $T^{+}_{cc}$ and
$X(3872)$, respectively. The resonant state in (b) is related to the
$X(3940)$. The y-axis of the eigenvalues in CSM method corresponds
to the half of the decay width. }\label{fig:csm}
\end{center}
\end{figure*}

The complex scaling method (CSM) is used to identify the bound and
resonant states. In CSM, the radius and momentum will rotate with an
angle $\theta$, $\boldsymbol{r} \rightarrow \boldsymbol{r} e^{i
\theta}, \quad \boldsymbol{q} \rightarrow \boldsymbol{q} e^{-i
\theta}$. With the varying $\theta$, the scattering states will
rotate with $2\theta$, while the bound and resonant states are
independent of the $\theta$ and will stay stable. More technical
details can be found in
Refs.~\cite{Aoyama:2006,Myo:2014ypa,moiseyev1998quantum}.
The complex eigenvalues for the $DD^{*}$ system ($T_{cc}$), the
$D\bar D^{(*)}$ system with $J^{PC}=1^{--}$ ($\psi(3770)$), and
$D\bar D^{*}$ system $J^{PC}=1^{++}$ ($X(3872)$) with the fixed
cutoff $\Lambda=1.0$ GeV
 are displayed Fig. \ref{fig:csm}.

\bibliography{TccX.bib}

\end{document}